\documentclass[prd,aps]{revtex4}

\usepackage{epsfig}

\begin{document}
\title{Quick and dirty methods for studying black-hole resonances.}
\author{Kostas Glampedakis}
\affiliation{ Department of Physics and Astronomy, Cardiff University 
P.O. Box 913, Cardiff, CF24 3YB, UK\footnote{Presently performing national
service with the Greek army.}} 

\author{Nils Andersson }
\affiliation{ Department of Mathematics, University of Southampton, 
Southampton SO17 1BJ, UK}

\date{\today}

\begin{abstract}
We discuss simple integration methods for the calculation of rotating black hole
scattering resonances both in the complex frequency plane (quasinormal modes) 
and the complex angular momentum plane (Regge poles). Our numerical schemes
are based on variations of ``phase-amplitude'' methods. In particular, we 
discuss the  Pr\"{u}fer transformation, where the original (frequency domain) Teukolsky 
wave equation is replaced by a pair of first-order non-linear equations governing the 
introduced phase functions. Numerical integration of these equations, performed along 
the real $r_{\ast}$-axis (where $r_{\ast}$ denotes the usual tortoise radial 
coordinate), or along rotated contours in the complex $r_{\ast}$-plane, provides 
the required 
${\cal S}$-matrix element (the ratio 
of amplitudes of the outgoing and ingoing waves at infinity). M\"uller's algorithm is 
then employed to conduct searches in the complex plane for the poles of this quantity 
(which are, by definition, the desired resonances). We have tested this method
by verifying known results for 
Schwarzschild quasinormal modes and Regge poles, and provide new results for the 
 Kerr black hole problem. We also describe a new 
method for estimating the ``excitation coefficients'' for quasinormal modes. 
The method is applied to scalar waves moving in the Kerr geometry, and the 
obtained results shed light on the long-lived 
quasinormal modes that exist for black holes rotating near the 
extreme Kerr limit.  
\end{abstract}
\maketitle


\section{Introduction}

During the last few decades it has become clear that resonant phenomena play
an important role in black hole physics \cite{novikov}. Numerical experiments, 
pioneered by Vishveshwara \cite{vishu} more than thirty years ago, 
and continued since then by many other authors, clearly show the dominance of the so-called 
quasinormal mode (QNM) oscillations in most dynamical processes involving black holes. 
The importance of this phenomenon is further emphasised by the possibility that future 
observations of QNM ``ringing'' following, say, the  merger of two black holes by the 
network of interferometric gravitational wave detectors  may provide 
direct identification of black holes. This exciting prospect has motivated a 
large number of investigations into the nature of the QNMs. An exhaustive 
list of papers on the subject can be found in 
Refs.~\cite{kokkotas,nollert_rev}.

The spectrum of QNM frequencies has been studied extensively in the literature, 
especially for the Schwarzschild black hole. It is well-known that the actual 
mode calculation is a non-trivial task. Recall that any physically 
acceptable solution to the relevant (homogeneous) wave equation takes the 
form of a purely ``ingoing'' wave $\sim \exp(-i\omega r_{\ast})$ 
(where $r_{\ast}$ is the tortoise radial coordinate) at the event horizon and 
a mixture of ``ingoing/outgoing''  waves 
$ \sim A_{\rm in} \exp(-i\omega r_{\ast}) + A_{\rm out} \exp(i\omega r_{\ast}) $  
far away from the black hole (such solutions exist because the effective curvature 
potential falls off asymptotically towards both the horizon and infinity, 
i.e. when $r_\ast\to\pm \infty$). 
A QNM solution corresponds to purely outgoing waves at infinity.
In other words, to search for a QNM one must first calculate 
the asymptotic amplitudes $A_{\rm out}$ and $A_{\rm in}$, and then
try to find a frequency that satisfies the required QNM boundary condition
$A_{\rm in}=0$. 
However, direct numerical integration of the wave equation leads to serious numerical
problems. Since the QNMs are expected to be damped due to gravitational-wave emission
the corresponding solutions must grow exponentially as one approaches infinity and 
the horizon (on a $t$-constant hypersurface).  In order to 
identify a QNM one must be able to filter out the solution that decays exponentially 
at infinity. In other words, the numerical calculation requires exponential precision.

Since the first  ``direct'' calculation of the slowest damped Schwarszchild QNMs  
by Chandrasekhar and Detweiler \cite{chandra1} various  alternative schemes have been 
devised. Detweiler \cite{detweiler1} was able to determine the fundamental Kerr QNM 
with adequate accuracy by examining the functional form of the asymptotic amplitudes 
along the real frequency axis.
Ferrari and Mashhoon \cite{valeria} exploited the correspondence between 
resonant and bound states of the inverted potential
to obtain approximate black hole modes.  WKB techniques 
were successfully used to calculate the first few least damped modes by 
Schutz and Will \cite{schutz} and several subsequent authors 
\cite{iyer1,iyer2,iyer3,guinn}.
High precision QNMs for both Schwarzschild and Kerr black holes
were first calculated by Leaver \cite{leaver1,leaver2}
by means of a continued fraction approach. His results for the Schwarzschild case
were later verified (and 
extended)  by Andersson \cite{na2} via the phase-amplitude formalism 
and  Nollert and Schmidt \cite{nollert1} who used a Laplace transform/analytic 
continuation method. Moreover, the semi-analytic phase-integral 
formalism \cite{froman} has proved to be a valuable tool for the calculation 
and interpretation of QNMs \cite{pi}.

More recently, Onozawa \cite{onozawa1} extended Leaver's calculation of the QNMs of 
rotating black holes and studied the behaviour of the mode frequencies as the rotation 
parameter is varied. The ``long-lived'' QNMs of a near extreme Kerr black hole 
were discussed by the present authors \cite{kg1}, using analytic expressions for 
the asymptotic amplitudes derived by Press and Teukolsky \cite{press}. 
That study complemented a calculation by Detweiler who was the first to use these 
amplitudes to derive a QNM-condition for near extreme rotation, and moreover, solved 
it analytically for the extreme case \cite{detweiler1}. Finally, Berti 
and Kokkotas \cite{bk03} have used the continued fraction approach to study the 
behaviour of the high overtone QNMs of a Kerr black hole. Their study 
unveils a complicated, and quite puzzling, structure.  

In the last few months 
there has been renewed interest in the spectrum of black hole QNMs. Motivated by the 
suggestion that the asymptotic behaviour of the high-overtone QNM frequencies 
\cite{noll,nahigh} may be related to the so-called Barbero-Immirzi parameter which is relevant 
for loop quantum gravity \cite{dreyer,motl}, there has been several attempts to 
approximate the high-frequency part of the spectrum. Some progress has been 
achieved by linking the high-overtone spectrum with the black hole's 
quantum properties \cite{hod}. Although we will not discuss this issue 
further in the present article, it is an interesting problem that need to be 
investigated in complete detail.  
 
The notion of Regge poles has (so far) hardly at all been discussed in the context of 
black holes. The Regge poles are close relatives to the QNMs \cite{jensen1}, and 
correspond to those complex values of the angular momentum $l$ for which the 
wavefunction is purely outgoing at infinity and purely ingoing at the horizon. 
The Regge pole calculation poses an easier problem than that of finding QNMs since 
the corresponding wavefunction is well behaved at infinity (and the horizon) for 
complex $l$ (as long as the frequency is kept real). In quantum  theory, the complex 
angular momentum (CAM) formalism (in which Regge poles play a central role) provides 
an alternative (and usually more transparent) description of scattering problems 
\cite{nussenzveig}. Some years ago, one of us applied the CAM theory to the study of 
scattering of massless scalar waves by a Schwarzschild black hole \cite{cam1,cam2}. 
In that study, Regge poles and the corresponding residues were calculated via 
the phase-integral method.  The CAM theory has also been used by Chandrasekhar and 
Ferrari \cite{chandra2} in an investigation of axisymmetric pulsations of a relativistic 
star.   

In this work we discuss  numerical methods for calculating both QNMs and Regge poles.
The key motivation for our study was the desire to devise methods that were easy to 
implement (hence ``quick and dirty'') and yet able to produce numerically accurate 
results. To achieve this goal we express the Teukolsky wave function in terms of 
suitable  phase-functions (see Section~III below). The integration of the resulting 
first-order equations readily yields the desired ratio of asymptotic amplitudes. 
In order to retain sufficient numerical precision to be able to identify the 
ingoing wave at infinity for complex frequencies, while ensuring that the  
phase-functions remain smooth and well behaved,  the integration is generally performed
along straight lines in the complex $r_{\ast}$ plane. 
The slope of the integration contour is chosen in such a way that 
the relative magnitude of the out- and ingoing wave solutions is preserved both 
near the horizon and at infinity. The end result is a multi-purpose scheme that  
(with minor modifications) facilitates the calculation of QNMs and Regge poles 
as well as scattering phase-shifts \cite{kg2}.

The remainder of the Paper is organized as follows: In Section II we briefly 
discuss the formal definition and physical interpretation of black hole resonances. 
Our numerical approaches to the problem are described in full detail in Section~III. 
Section~IV contains  numerical results for QNMs (Section IVA) and 
Regge poles (Section IVB). Section~V provides a summary and some final comments.  
Geometric units $G=c=1$ are adopted throughout the paper.


\section{Black hole resonances: definition and physical meaning}

The notion of scattering resonances is probably most familiar from quantum theory 
\cite{taylor}. The fact that it is also relevant in black hole physics (in terms of 
QNMs and Regge poles) is not particularly surprising given that linear perturbations 
of black hole spacetimes are governed by Schr\"{o}dinger type equations. 
The corresponding effective 
potential barriers are a combination of the black hole's gravitational pull and a 
repulsive centrifugal term \cite{chandrabook}. This means that one can formulate a 
scattering problem for black holes and investigate  phenomena analogous to those
in the quantum case. For simplicity we will mainly consider the case 
of scalar waves in this paper. This is not a severe restriction, however, since
more ``realistic'' cases, eg. gravitational perturbations, can 
be dealt with in exactly the same way.   
 
After adopting the Boyer-Lindquist coordinate frame, and  separating
the angular dependence in the standard way, 
we arrive at the Teukolsky equation \cite{teuk},
which governs the radial component $\psi(r)$ of a scalar 
perturbation in a Kerr black hole spacetime,
\begin{equation}
\frac{d^2 \psi}{dr^2_{\ast}} + Q_{l m}(r_{\ast},\omega,a) \psi =0
\label{schro}
\end{equation}
We have also assumed a harmonic time dependence $ e^{-i\omega t}$.
The tortoise coordinate $r_{\ast}$ is defined as
\begin{equation}
\frac{dr_{\ast}}{dr}= \frac{r^2 + a^2}{\Delta}
\end{equation}
where $\Delta= (r-r_{+})(r-r_{-})$ and $r_{\pm}= M \pm (M^2 -a^2)^{1/2}$
denote the black hole's inner  (Cauchy) and outer (event) horizon.  Explicitly we have
\begin{equation}
r_{\ast}= r + \frac{2Mr_{+}}{r_{+}- r_{-}} \ln \left ( \frac{r}{r_{+}} -1
\right ) - \frac{2Mr_{-}}{r_{+} -r_{-}} \ln \left ( \frac{r}{r_{-}} -1
\right )
\label{tort}
\end{equation}

The effective potential $Q_{l m}$ is given by
\begin{equation}
Q_{l m}(r_{\ast}, \omega,a)= \frac{ K^2 + (2am\omega -a^2 \omega^2 -E_{l m}) \Delta} 
{(r^2 + a^2 )^2} -\frac{dG}{dr_{\ast}} -G^2
\label{pot}
\end{equation}
with $ K= (r^2 +a^2)\omega -am$ and $G= r\Delta/(r^2 + a^2)^2 $.
It has the following asymptotic behaviour
\begin{equation}
Q_{l m} \sim \left\{ \begin{array}{ll} k^2
\quad \mbox{as } r\to r_+ \ (r_\ast\to-\infty)\ , \\ \omega^2  \quad
\mbox{as } r\to +\infty \ (r_\ast\to\infty)\ , \end{array} \right.
\label{asympt1}
\end{equation}
where $ k= \omega - m\omega_{+} $ with $\omega_{+}= a/2Mr_{+} $ denoting
the angular frequency of the event horizon. The eigenvalue of the relevant angular
equation is denoted by $E_{l m}$, cf. \cite{teuk}. 

As a result of (\ref{asympt1}), the solution 
of (\ref{schro}) will (to leading order) be a linear combination of simple 
exponentials as $ r_{\ast} \to \pm \infty$. A physically acceptable solution 
describes a wave propagating towards the horizon at $r_{\ast} \to -\infty$, 
i.e. a purely ``ingoing'' solution. We can normalize this solution according to
\begin{equation}
\psi^{in} \sim \left\{ \begin{array}{ll}
e^{-ikr_{\ast}} \quad \mbox{as } r_{\ast} \to -\infty \ , \\ 
A_{\rm in} e^{-i\omega r_{\ast}} + A_{\rm out} e^{i\omega r_{\ast}} \quad\mbox{as } 
r_{\ast} \to +\infty \ .
\end{array} \right.
\label{in}
\end{equation}
Another useful solution is  purely ``outgoing'' at infinity,
\begin{equation}
\psi^{up} \sim e^{i\omega r_{\ast}}
\label{up}
\end{equation}
The Wronskian of any two linearly independent solutions of (\ref{schro})
is constant. Using the two solutions above we easily get
\begin{equation}
W(\psi^{in},\psi^{up}) = 2i\omega A_{\rm in}
\label{wronsk}
\end{equation}
Finally, we define the ${\cal S}$-matrix, a quantity familiar from scattering theory 
\cite{taylor}, as the ratio
\begin{equation}
{\cal S} (\omega,l) \equiv (-1)^{\ell +1} \frac{A^{\rm out}}{A^{\rm in}}
 =   \exp(2i\delta_{l})
\label{smatrix}
\end{equation}
where $\delta_l$ defines the scattering phase-shifts.

One can define two classes of ``resonances'' that correspond to purely ingoing waves 
at the horizon and purely outgoing waves at infinity. Loosely speaking, 
such solutions are ``characteristic'' of the black hole spacetime since they 
do not depend on waves coming in from infinity.
The first class of resonances are the QNMs. They correspond to complex 
frequency  poles $\omega_n$ of ${\cal S}$ (for integer $l$).
The QNM frequencies (obviously) correspond to solutions 
such that the Wronskian (\ref{wronsk}) vanishes. 
The second possible class of resonances are the Regge poles. To identify these one 
allows the angular momentum ($l$) to assume complex values (while keeping $\omega$ real)
and searches for poles $l_n$ of ${\cal S}$. The close connection between these classes 
of solutions has recently been discussed by D\'ecanini, Folacci 
and Jensen~\cite{jensen1}.

It is well-known that the numerical determination of QNMs requires considerable care.
Since we are discussing 
(stable) damped modes, the QNM frequencies will have negative imaginary parts. 
It follows then, that each QNM eigenfunction  
$ e^{-i\omega_n t} \psi^{\rm in}(\omega_n, r_{\ast}) $  will grow exponentially 
both towards infinity and at the horizon. Since one must be able to filter out the 
ingoing-wave solution at infinity in order to identify a QNM, the calculation 
requires exponential precision. 
It should be pointed out that this problem is not physical: It arises because we
are viewing only a part of the overall problem, as is clearly shown in discussions of 
the initial-value problem for perturbed black holes 
\cite{nollert1,kokkotas,naprd}. Those studies reveal the true physical meaning of QNMs:
They describe part of the late time behaviour of the perturbing field
as observed at some fixed distance from the black hole \cite{nollert1}.
They can  be viewed as arising as a part of the field is temporarily ``trapped'' 
near the scattering center (the black hole) and leaks out to infinity. 
The same picture arises in the study of QNMs of much simpler systems such as 
static objects scattering waves in flat spacetime \cite{jensen2,lax}.

The QNMs  arise in scattering situations that can be thought of as 
``time dependent'' (strictly speaking, the  evolution and scattering  of 
initial data with compact support). Alternatively, one  might consider a 
time-independent  scattering ``experiment''. Namely, the scattering of monochromatic 
plane waves (where the time dependence is trivially given by $\sim \exp(-i\omega t)$
for any given $\omega$). 
In this case, the most important ``observable'' quantity is the differential cross section, 
which is a measure of the scatterer's size as seen by waves scattered into a certain 
solid angle \cite{taylor}. The field at infinity consists of the incoming plane wave 
plus a scattered piece, and can be written
\begin{equation}
\psi (r_{\ast} \to + \infty) \sim \psi_{\rm plane} + f(\theta) 
\frac{e^{i\omega r_{\ast}}}{r}
\label{scat_bc}
\end{equation}
where we have assumed, for simplicity, axisymmetric scattering (for a treatment
of non-axisymmetric scattering by rotating black holes, see \cite{kg2}). 
The cross section can  be expressed in terms of the (complex-valued) scattering
amplitude $f(\theta)$,
\begin{equation}
\frac{d\sigma}{d\Omega}= |f(\theta)|^2
\label{cs}
\end{equation}
For scalar waves scattered off a Schwarzschild black hole one finds that \cite{matzner},
\begin{equation}
f(\theta)= \frac{1}{2i\omega} \sum_{l=0}^{\infty} (2l +1) 
( {\cal S}_{l} -1) P_{l}(\cos\theta)
\label{scat_amplit}
\end{equation}
from which we see that all the relevant scattering information is encoded in
the ${\cal S}$-matrix. 

The angular scattering problem is usually studied in terms of the phase-shifts 
$\delta_l$. However, it is not unusual for the sum in (\ref{scat_amplit}) to be  
slowly converging  (black-hole scattering being a example, although this difficulty 
can to some extent be avoided following the prescription in \cite{handler}). 
In many situations it is therefore desirable to have a different representation for 
the scattering amplitude. The most celebrated alternative  is the CAM approach 
\cite{nussenzveig}. By means of a Watson-Sommerfeld transformation the sum over $l$ 
in (\ref{scat_amplit}) can be rewritten as an integral along a contour in the complex 
$l$-plane \cite{cam1}. This integral naturally splits into two terms,
\begin{equation}
f(\theta)= f_{P}(\theta) + f_{B}(\theta)
\label{split}
\end{equation}
The first term is a sum over the residues of the 
poles $l_n$ of ${\cal S}$; the  Regge poles  \cite{taylor}. 
It is explicitly given by,
\begin{equation}
f_{P}(\theta)= -\frac{i\pi}{\omega} \sum_{n=0}^{\infty} 
\frac{r_n (l_n + 1/2)}{\cos\pi(l_n + 1/2)} P_{l_n}(-\cos\theta)
\label{fp}
\end{equation}
where $r_n$ is the residue of each Regge pole, i.e. is defined by
\begin{equation}
{\cal S} \approx { r_n \over l-l_n} 
\end{equation}
in the vicinity of the $n^{\rm th}$ pole, 
 and $P_{l}$ is a Legendre function of complex degree.
The second term in (\ref{split}) is a  ``background integral'', 
which is known to be less relevant as far as diffraction effects are  concerned. 
As was demonstrated in \cite{cam2},  only the 
first few poles are required for determining the scattering cross section 
for angles away from the forward direction (the direction of incoming plane wave). 
Despite the fact that the problem of  scattering by rotating black holes
has not yet been studied within the CAM framework, it is natural to  define Kerr black hole
Regge poles as singularities 
of ${\cal S}$ in the complex $l$ plane while keeping $\omega$ real and $m$ an integer. 

The physical interpretation of Regge states is discussed in standard textbooks
\cite{nussenzveig,taylor}, cf. also the discussion of  scattering of plane waves 
by a Schwarzschild black hole in \cite{cam2}. 
By approximating the Legendre function for
$|l_n + 1/2| \gg 1$ and $\theta$ not too close to $0$ or $\pi$ one can show that
\begin{equation}
P_{l_n}(-\cos\theta) \approx \left[ \frac{1}{2\pi(l_n +1/2)
\sin\theta} \right]^{1/2} \{ e^{i(l_n + 1/2)(\pi -\theta) -i\pi/4}
+ e^{-i(l_n + 1/2)(\pi -\theta) + i\pi/4} \}
\label{approxP}
\end{equation}
From this formula it follows that each Regge pole can be interpreted 
as a combination of two ``surface'' waves traveling around the black hole 
in opposite directions. These waves decay exponentially as they propagate, 
with an ``angular life'' $\sim 1/\mbox{Im}(l_n + 1/2)$. The real part of 
the Regge pole is associated with the distance $R_n$ from the black hole 
at which the angular decay occurs, Re~$(l_n +1/2) \approx \omega R_n$ (note 
that this ``localization principle'' is formally valid only for $l \gg 1$ \cite{taylor}). 
As can be verified using the numerical results presented in Section VI, the 
value of $R_n$ lies very close to the critical impact parameter for 
null geodesics (which is $b_{\rm c}= 3\sqrt{3}M$  for a Schwarzschild black hole). 
This suggests that (as one would have expected) Regge states 
describe waves temporarily trapped in the vicinity of the unstable photon orbit, 
leaking to infinity and to the horizon, as they travel around the black hole.

\section{Two ``quick and dirty'' schemes}

In this Section we will discuss two closely related methods for 
determining black-hole resonances. Both methods are based on the idea
of integrating slowly varying phase-functions rather than the, possibly rapidly,
oscillating wavefunction itself. The representation of the solutions 
to Schr\"odinger type equations in terms of phase-functions was 
first discussed by Milne \cite{milne} and, a long time later, by 
Newman and Thorson \cite{newman}. The second key ingredient in our schemes
is the use of complex values of the radial coordinate.  
By analytically continuing the resonant solutions into the complex coordinate plane 
one can suppress the divergences associated with (in particular) the 
QNM boundary conditions. 
In essence, our methods are close relatives to
the phase-amplitude method that was developed by one of us \cite{na2} 
to study of scattering resonances. The main 
difference is that the methods we describe here make use of 
much simpler integration contours, and are consequently much easier 
to implement.


\subsection{The Pr\"{u}fer method}

The family of the so-called Pr\"{u}fer transformations has been widely used
in the numerical analysis of Sturm-Liouville problems \cite{pryce}.
It has also been applied to the calculation of resonances and phase-shifts 
in standard quantum scattering problems, see for example Refs.~\cite{pajunen1,pajunen2}.

Consider a Schr\"{o}dinger-type equation, where either the frequency $\omega$ or
the angular momentum $l$ are allowed to take complex values,
\begin{equation}
\frac{d^2 \psi}{dx^2} + Q(x, l, \omega) \psi=0
\label{sle}
\end{equation}
We consider cases where the radial coordinate $x$  spans the entire
real axis (in the black hole case $x$ corresponds to the tortoise 
coordinate $r_{\ast}$), and the effective potential $Q$ is 
a single barrier with asymptotic behaviour
\begin{equation}
Q \sim \left\{ \begin{array}{ll}
q^2 \quad \mbox{as } x  \to -\infty  \\ 
\omega^2 \quad\mbox{as } x \to +\infty 
\end{array} \right.
\label{Qapprox}
\end{equation} 
where $q=q(\omega)$. It follows then, that the function 
$\psi(x,l,\omega)$ will 
be a linear combination of exponentials as $x \to \pm \infty$. Following the 
discussion in the previous Section, let us assume that we are interested in 
a solution to (\ref{sle}) which is purely ``ingoing''  as 
$x \to - \infty$ and mixed ingoing/outgoing as $x \to + \infty$. That is, we want to 
determine the ``in'' solution, see eqn. (\ref{in}), which can be written  
\begin{equation}
\psi \sim \left\{ \begin{array}{ll}
e^{-iqx} \quad \mbox{as } x  \to -\infty \ , \\ 
B \sin[\omega x + \zeta ] \quad\mbox{as } x \to +\infty \ ,
\end{array} \right.
\end{equation}
where $\zeta$ and $B$ are some complex constants. 

We can write the 
exact solution of (\ref{sle}) in the form 
\begin{equation}
\psi(x,l,\omega)= \exp[\int P(x,l,\omega) dx ] 
\end{equation}
where $P$ is the logarithmic derivative of $\psi$ (a prime denotes derivative 
with respect to $x$),
\begin{equation}
\frac{\psi^{\prime}}{\psi}= P
\label{logd}
\end{equation}
This is our first phase-function. Obviously, it should obey the boundary 
condition $P \to -iq$ for $x \to -\infty$.
Alternatively, we can express the function $\psi$ and its derivative via the transformation, 
\begin{eqnarray}
\psi(x,l,\omega) &=& B \sin[\omega x +\tilde{P}(x,l,\omega)]
\label{prufer} 
\\
\psi^{\prime}(x,l,\omega)&=& B \omega \cos[\omega x +
\tilde{P}(x,l,\omega) ]
\label{prufderiv}
\end{eqnarray}
where $\tilde{P}$ is the Pr\"{u}fer phase function (which should
have $\zeta$ as 
its limiting value as $x \to + \infty$). 
    
The two phase functions satisfy non-linear
first order differential equations;
\begin{equation}
\frac{dP}{dx} + P^2 + Q = 0
\label{P1}
\end{equation}
\begin{equation}
\frac{d\tilde{P}}{dx} + \left ( \omega -\frac{Q}{\omega} \right )
\sin^2(\omega x + \tilde{P})= 0
\label{P2}
\end{equation}
and they are directly related by
\begin{equation}
\tilde{P} = -\omega x + \frac{1}{2i} \log\left[ \frac{iP -\omega}
{iP +\omega} \right ]
\end{equation}
or
\begin{equation}
P = \omega \cot( \omega x + \tilde{P} )
\end{equation}

A major advantage of using phase-functions can be seen immediately
from Eqns. (\ref{P1}) and (\ref{P2}), assuming for the moment that
we are dealing with real $\omega$ and $l$. 
Using (\ref{Qapprox}), we see that (\ref{P1}) and (\ref{P2}) lead to
\begin{eqnarray}
\frac{dP}{dx} &\approx & 0 \quad \mbox{as} \quad x \to -\infty
\\
\nonumber \\
\frac{d\tilde{P}}{dx} &\approx & 0 \quad  \mbox{as} \quad x \to +\infty 
\end{eqnarray}
Hence, the functions $P$ and $\tilde{P}$ are slowly varying for large negative and positive 
values of $x$, respectively. This means that the oscillatory behaviour of the 
original wavefunction in those regimes 
[which would  cause the build up of unnecessary numerical 
errors in a direct integration 
of (\ref{sle})] has been effectively removed.

Finally, knowledge of the $\tilde{P}$ phase-function at $x \to +\infty$ 
provides us with the ${\cal S}$-matrix,
\begin{equation}
{\cal S}(l,\omega)= (-1)^{l+1} e^{2i\zeta}
\label{Smatrix}
\end{equation}

Obviously, the described method can be applied to the black-hole
problem with $ x \equiv r_{\ast} $, $ q \equiv k $ and (\ref{sle}) becoming the Teukolsky 
radial equation (\ref{schro}). 

\begin{figure}[tbh]
\centerline{\epsfysize=8cm \epsfbox{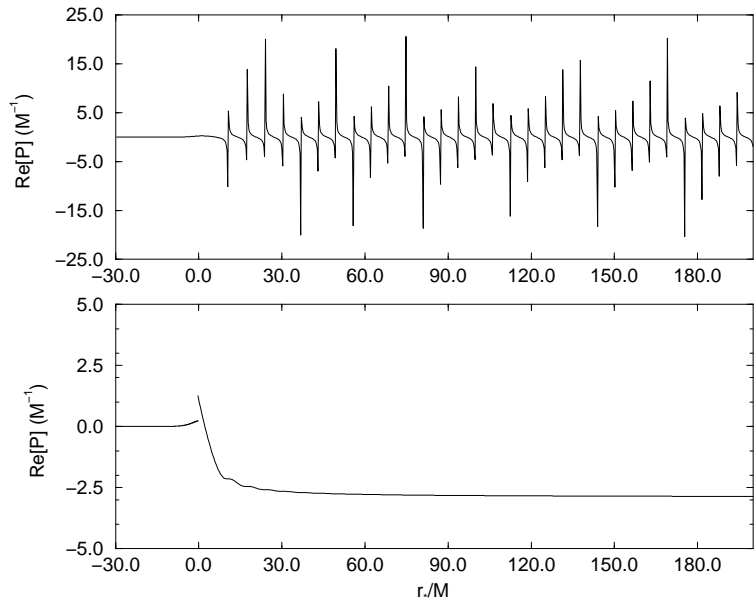}}
\caption{The Pr\"ufer phase-functions (only real parts shown) 
resulting from a typical integration of eqns (\ref{P1}) and (\ref{P2}) with parameters 
$l=m=2, \omega M=0.5, a=0.5M$. Top panel: the integration of the $P$ function 
was continued beyond the potential peak (located close to $r_{\ast}= 0 $). Beyond this
point its 
behaviour changes dramatically as a result of the Stokes phenomenon.
As is clear from the graph, the derivative of  $P$ takes particularly large values at a 
series of points beyond $ r_{\ast}=0 $. This may signal the existence of poles of the 
phase function close to the real $r_{\ast}$ axis. That this behaviour could be expected, 
if one fails to account for the Stokes phenomenon, is clear from the equations given 
in Appendix 1 of \cite{cam3}. 
Bottom panel: as soon as the peak region is reached, we shift to the 
$\tilde{P}$ function (the matching is performed at the point where the curve ``breaks'' 
in the figure) which is much better behaved, basically because it describes both 
ingoing and outgoing waves, see eqn. (\ref{prufer}). Tests show that the
result does not depend on the choice of 
matching  point, as long as it remains close to the peak of the effective black-hole
potential.}
\label{stok}
\end{figure}

At the practical level, the numerical integration for the $P$ function 
shoud be initiated at a point $r_\ast^H$ (say), 
close to the event horizon. 
Assuming that $Q$ is a slowly varying function for large
$|r_\ast|$, which is certainly the case in the applications under
consideration, we can initiate the integration of (\ref{P1}) 
from the lowest order WKB approximation
\begin{equation}
P \approx  \pm iQ^{1/2} \qquad \mbox{ as } |r_\ast|\to
\infty
\end{equation}
(alternative approximations, eg. a power series
expansion, will work equally well).
In this way we can readily find a solution that corresponds to the
(analytically continued) ingoing wave boundary condition at the
horizon. This solution is then
integrated towards the  matching point $r_{\ast}^{\rm ma}$.
As the integration 
reaches the vicinity of the potential peak (the local minimum of $Q$ in the Kerr case) 
we switch 
to the $\tilde{P}$ function, and then carry on the integration to ``infinity''. 
It is mandatory to change phase-function if one wants to ensure that the 
integrated function remains smooth and well behaved. This is due to the fact 
that the phase-functions undergo a drastic change in their behaviour (they 
become rapidly oscillating and may even have singularities) as they are extended 
through the region of the potential peak. This is a manifestation of the so-called 
Stokes phenomenon --- the ``switching on'' of small exponentials in the solution 
to an equation of form (\ref{sle}), well known from WKB-and phase-integral theory 
\cite{berry,froman} (see also the discussion in Appendix 1 of \cite{cam3}).
This situation is demonstrated in Fig.~\ref{stok} where we plot the functions 
$P(r_\ast)$ and $\tilde{P}(r_\ast)$ for  typical parameters  
$l=m=2$, $\omega M= 0.5$, $a=0.5M$. 

The Pr\"{u}fer approach has already been applied in the context of scattering of
plane scalar waves by rotating black holes \cite{kg2}. In that study, where 
both $\omega$ and $l$ were assumed to be real, the use of the phase-functions 
described above facilitated 
the computation of accurate high frequency phase-shifts (which 
would be problematic if the original wave equation were to be integrated \cite{handler}). 
We can follow exactly the same procedure when $l$ is allowed to acquire complex values 
(as long as $\omega$ remains real). In this way we have calculated 
the first few Regge poles of both
Schwarzschild and Kerr black holes. The numerical results will be 
discussed in the following Section.

\begin{figure}[tbh]
\centerline{\epsfysize=5cm  \epsfbox{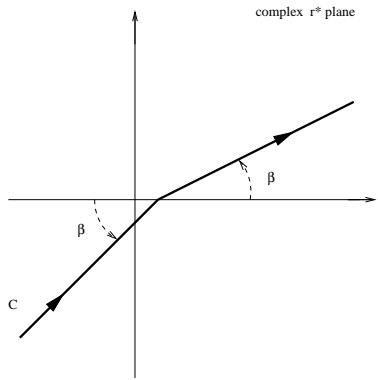}}
\caption{The integration path $C$ in the complex $r_{\ast}$-plane used to 
calculate QNMs. The slope $\beta$ is chosen in such a way that 
divergent/decaying solutions at infinity and at the horizon are avoided.} 
\label{contourfig}
\end{figure}

Complications arise when $\omega$ becomes complex while $l$ 
is kept real (integer). From Eqn.~(\ref{P2}) we see that the term which represents 
an ingoing (outgoing) wave becomes exponentially small (large) at large $x$ 
(for Im~$\omega <0$). A similar situation appears close to the horizon. 
Therefore, we are still facing one of the problems associated with a 
direct integration of the wave equation. An efficient and elegant 
way to avoid this difficulty is to consider the problem in the complex  $r_{\ast}$-plane. 
The success of this approach has been repeatedly demonstrated with 
the application of the phase-integral/amplitude formalisms to the calculation 
of black-hole QNMs \cite{na2,na6,pi}. The main idea is to integrate our 
equations not along the real axis, but instead use a suitably chosen contour $C$ 
such that the ingoing/outgoing wave solutions, eg. $e^{\pm i\omega r_{\ast}}$, have comparable 
asymptotic magnitudes (of order  unity) as $r_\ast\to \infty$. 
The simplest choice of integration contour is a 
straight line with slope 
$-$Im~$\omega/$ Re~$\omega$, see Fig~\ref{contourfig}. Similarly, a line with 
slope $-$Im $ k/$ Re $k$ will guarantee purely oscillatory behaviour 
near the event horizon. Such paths are parallel 
(for $|r_{\ast}| \to \infty$) to the so-called anti-Stokes lines which 
play a crucial role in WKB and phase-integral theory \cite{froman,na2}. 
The anti-Stokes lines are curves along which $\int Q^{1/2} dr_{\ast} $ is purely
real and as a consequence the WKB 
solution of (\ref{schro}) is purely oscillatory. 
It should be noted that a similar integration contour has been used in a 
calculation of relativistic stellar pulsation modes \cite{na4}. 

We use a parameterization
\begin{equation}
r_{\ast}= r_{\ast}^{\rm ma} + \rho e^{i\beta}
\label{contour}
\end{equation}
with $ -\infty < \rho < +\infty$ and $\beta$ given by,
\begin{equation}
\beta = \left\{ \begin{array}{ll} -Arg(\omega)
\quad \mbox{for } \rho > 0 \ , \\ -Arg(k)  \quad 
\mbox{for } \rho < 0 \ , \end{array} \right.
\label{slope}
\end{equation}
The matching point  $r_{\ast}^{\rm ma}$  is chosen to lie on the real 
axis ( $\rho=0 $) and some experimentation shows that reliable results 
are obtained as long as the matching is performed near the 
peak of the curvature potential (the extremum of $Q$).
    
Rewriting Eqns. (\ref{P1}) and (\ref{P2}) in terms of $\rho$ we have
\begin{equation}
\frac{dP}{d\rho} + e^{-i\beta} \left ( P^2 + Q \right ) = 0
\label{P1c}
\end{equation}
\begin{equation}
\frac{d\tilde{P}}{d\rho} +  e^{-i\beta}\left ( \omega -\frac{Q}{\omega} \right )
\sin^2(\tilde{P} + |\omega|\rho + \omega r_{\ast}^{\rm ma} )= 0
\label{P2c}
\end{equation}
The radial coordinate $r$ will also be assumed complex, 
and we determine it from
\begin{equation}
\frac{dr}{d\rho}= e^{-i\beta} \frac{\Delta(r)}{r^2 + a^2}
\label{tortc}
\end{equation}
With the help of (\ref{tort}) it is easy to see that
\begin{equation}
r \approx r_{+} ( 1 + e^{c r_{\ast}})
\label{spiral}
\end{equation}
where $c= (r_{+} -r_{-})/2Mr_{+}$  as $\rho \to -\infty $, and that 
$r \approx r_{\ast} $ as $\rho \to +\infty $. 
Hence, the proposed integration contour corresponds to a 
straight line at large distances in the $r$-plane, and to a clockwise spiral 
as the horizon is approached. This strongly resembles 
the more complicated integration contour used in the 
phase-amplitude calculation of QNMs, see figures 1 and 2   
in \cite{na2}.  

Not surprisingly, 
some care must be taken in implementing complex integration contours
in black-hole problems. For example, we need to understand the analytic 
properties of the potential $Q$. From Eqn.~(\ref{pot}) we deduce
that this quantity will have poles at $r=\pm ia$. 
Since the function $r_{\ast}(r)$ 
is multivalued, 
there will be a series of poles in the left half of the 
$r_{\ast}$-plane,  symmetrically located with respect to the real axis. 
All of these poles have the same real part 
(which is zero for the non-rotating case), while their imaginary 
parts satisfy Im~$r_{\ast} \gg \mbox{Re}~r_{\ast}$. 
In practice, the presence of these singularities provides an upper 
limit to  the complex-rotation angle. This is easiest illustrated by 
the Schwarzschild case. According to (\ref{tort})
the point $r=0$ translates into
\begin{equation}
r_{\ast}(0)= 2M \log(-1)= \pm 2M i\pi (1 + 2k) \quad \mbox{with}
\quad k= 0,1,2,...
\label{singu}\end{equation}
It is easy to understand that 
one should not rotate the integration contour beyond this
angle.
     
Another important issue that deserves some discussion concerns 
the boundary conditions in the complex plane. When we are solving the system of 
phase equations (\ref{P1c}) and (\ref{P2c}) we are actually 
imposing the QNM boundary condition on the contour $C$ at $\rho \to \pm \infty$. 
However, the ``true'' boundary conditions are to be imposed on the real axis 
($r_{\ast} \to \pm \infty$). It is perfectly natural to wonder if we are, in 
some way, altering the problem by using the contour $C$. This question 
 has been discussed in the context of 
WKB and phase-integral studies (and the complex $r$-plane) 
\cite{araujo,froman}. It was concluded that it is safe to analytically 
continue the boundary condition in the proposed way.


\subsection{QNM excitation coefficients}

We have used the Pr\"ufer transformation 
scheme described above to calculate both QNMs and Regge
poles for scalar waves in the geometry of a rotating black hole. 
The method is easily extended to other classes of perturbing fields, 
the most interesting of which is likely to be that of gravitational 
perturbations. As will be discussed below the method is somewhat
limited for extremely rapidly damped QNMs, and hence it cannot 
be used to investigate the asymptotic behaviour of 
the mode spectrum \cite{motl}. However, from an astrophysical 
point of view the slowly damped modes are the most important
and for these our scheme works very well, indeed. 

In the case of QNMs one might also want to assess to what extent the
modes are excited given a certain initial perturbation. 
As discussed in \cite{naprd} the answer can be quantified in an initial-data
independent way in terms 
of the ``excitation coefficient''
\begin{equation}
B_n = { A_{\rm out}(\omega_n) \over 2 \omega_n \alpha_n}
\end{equation} 
where we have assumed that 
$A_{\rm in} \approx \alpha_n (\omega-\omega_n)$ in the 
vicinity of the $n^{th}$ mode frequency.
While one can formulate expressions for this quantity within our
previous scheme, we have found that the precision of the actual 
numerical results is somewhat unsatisfactory. To remedy this we have 
devised an alternative scheme.
The purpose of the following discussion is to present this scheme, 
with the particular aim of achieving accurate results for the excitation
coefficients.

The first half of the calculation remains 
identical to the one discussed above, and we thus  
obtain a solution $P_L(r_{\ast}^{\rm ma})$ (say) which satisfies that boundary 
condition of ingoing waves at the event horizon. At
spatial infinity we now obtain the appropriate solution in a similar
way. Let us choose the phase of the square root in such a 
way that the solution corresponding to $P_+ \approx +
iQ^{1/2}(r_{\ast})$ has the desired outgoing-wave character. Once
this solution is extended to the matching point we have a second solution
$P_+(r_{\ast}^{\rm ma})$. Thus a resonant overall solution, a QNM, 
follows from the simple condition
\begin{equation}
P_+(r_{\ast}^{\rm ma}) = P_L(r_{\ast}^{\rm ma})
\end{equation}

Now we want to determine also the excitation coefficients. 
We can do this by considering the solution for a
general frequency, that contains both in- and outgoing wave at
infinity, rather than a specific QNM solution. 
Assume (as before) that $P_+ \approx +
iQ^{1/2}(r_\ast)$ corresponds to an outgoing wave solution, and
that  $P_- \approx - iQ^{1/2}(r_\ast)$ represents an
ingoing wave. This means that, after integrating these two
functions to the matching point the general
solution corresponds to the linear combination
\begin{equation}
\psi_R(r_\ast^{\rm ma}) = 
A \exp \left[ \int_{r_\ast^\infty}^{r_\ast^{\rm ma}} P_+
dr_\ast \right] + B \exp \left[ \int_{r_\ast^\infty}^{r_\ast^{\rm ma}} P_-
dr_\ast \right]
\end{equation}
where the integration is initiated at $r_\ast^\infty$.
We can write the logarithmic derivative of this function as
\begin{equation}
{ \psi_R^\prime \over \psi_R} = { P_+ + (B/A) P_-
\exp\left[ \int_{r_\ast^\infty}^{r_\ast^{\rm ma}} ( P_- -P_+) dr_\ast
\right] \over 1 + (B/A) \exp\left[  \int_{r_\ast^\infty}^{r_\ast^{\rm ma}} (
P_- -P_+) dr_\ast \right] }
\end{equation}

The condition for an overall solution at $r_\ast^{\rm ma}$ is that the
logarithmic derivative is continuous, i.e. that the above
expression is equal to $P_L(r_\ast^{\rm ma})$. This condition 
can be written
\begin{equation}
{B\over A} = { P_+ - P_L \over P_L - P_-}
\exp\left[  \int_{r_\ast^\infty}^{r_\ast^{\rm ma}} ( P_+ -P_-) dr_\ast
\right] \quad \mbox{ for } r_\ast = r_\ast^{\rm ma}
\end{equation}
Now recalling that $P_-$ corresponds to the ingoing wave
solution at infinity we know that a QNM must correspond to $B=0$,
and we see that the above condition is the same as the one we
derived previously.

To determine the excitation coefficient requires a little more
work. First we need to relate our amplitudes $A$ and $B$ to the
asymptotic amplitudes $A_{\rm out}$ and $A_{\rm in}$. Recall that
the asymptotic amplitudes are determined from
\begin{equation}
\psi \sim A_{\rm out} e^{i\omega r_\ast} +  A_{\rm in} e^{-i\omega
r_\ast} \quad \mbox{ as } r_\ast \to \infty
\end{equation}
We want to compare this expression to our solutions. Since we have
assumed that the WKB approximation is valid at the point where we
initiate integration ($r_\ast^\infty$) we can use
\begin{equation}
\int_{r_\ast^\infty}^{r_\ast} P_\pm dr_\ast \approx \pm i
\int_{r_\ast^\infty}^{r_\ast} Q^{1/2} dr_\ast = \pm i\omega (r_\ast -
r_\ast^\infty) \pm i\int_{r_\ast^\infty}^{r_\ast} \left( {Q-\omega^2 \over
Q^{1/2} + \omega} \right)  dr_\ast
\end{equation}
beyond $r_\ast^\infty$
(the last integral should be easy to evaluate). We then have
\begin{equation}
\psi_+ \approx A \exp \left[ \int_{r_\ast^\infty}^{r_\ast} P_+
dr_\ast \right] = A \exp \left[ i\omega (r_\ast - r_\ast^\infty)+
i\int_{r_\ast^\infty}^{r_\ast} \left( {Q-\omega^2 \over Q^{1/2} +
\omega} \right)  dr_\ast \right] \mbox{ for } |r_\ast| >
|r_\ast^\infty|
\end{equation}
And we can identify
\begin{equation}
A_{\rm out} = A \exp \left[-i \omega r_\ast^\infty+
i\int_{r_\ast^\infty}^\infty \left( {Q-\omega^2 \over Q^{1/2} + \omega}
\right)  dr_\ast \right]
\end{equation}
Similarly, we get
\begin{equation}
A_{\rm in} = B \exp \left[i \omega r_\ast^\infty -
i\int_{r_\ast^\infty}^\infty \left( {Q-\omega^2 \over Q^{1/2} + \omega}
\right)  dr_\ast \right]
\end{equation}

Collecting the above results, we have
\begin{equation}
{ A_{\rm in} \over A_{\rm out} } = { P_+ - P_L \over
P_L - P_-} \exp\left[ \int_{r_\ast^\infty}^{r_\ast^{\rm ma}}( P_+
-P_-) dr_\ast + 2i \omega r_\ast^\infty -
2i\int_{r_\ast^\infty}^\infty \left( {Q-\omega^2 \over Q^{1/2} +
\omega} \right)  dr_\ast \right]
\end{equation}
The final step in the derivation consists of taking the derivative
of this expression with respect to $\omega$ and evaluating it at a
QNM frequency. This immediately leads to
\begin{equation}
{\alpha_n \over A_{\rm out} } =   {1 \over P_L - P_-}
\left[ {d\over d\omega} (P_+ - P_L) \right] \exp\left[
\int_{r_\ast^\infty}^{r_\ast^{\rm ma}}( P_+ -P_-) dr_\ast + 2i \omega
r_\ast^\infty - 2i\int_{r_\ast^\infty}^\infty \left( {Q-\omega^2 \over
Q^{1/2} + \omega} \right)  dr_\ast \right]
\end{equation}
from which we can calculate the desired excitation
coefficient. 

To conclude, let us summarize the additional steps that are
required in a calculation of the excitation coefficient once a QNM
frequency has been determined.
First we need to evaluate the two integrals
$\int_{r_\ast^\infty}^{r_\ast^{\rm ma}}P_\pm dr_\ast$. This is best done by
introducing additional dependent variables, let us call them
$\Phi_\pm$, and then  integrating the equations
\begin{equation}
\Phi^\prime_\pm = P_\pm
\end{equation}
together with (\ref{P1}) from $r_\ast^\infty$ to $r_\ast^{\rm ma}$.
Secondly, we need the derivatives ${dP_+ \over d\omega} $
and ${dP_L \over d\omega} $. These quantities are also best 
calculated by
introducing additional variables, i.e. by integrating
\begin{equation}
{d \over dr_\ast} \left( {dP_\pm \over d\omega } \right) = - {dQ
\over d\omega} - 2P_\pm {dP_\pm \over d\omega}
\end{equation}
using appropriate initial data, i.e.
\begin{equation}
{dP_\pm \over d\omega} \approx \pm {i \over 2Q^{1/2}}{dQ \over
d\omega}
\end{equation}
at $r_\ast^\infty$.
Thirdly, we need to evaluate
\begin{equation}
\int_{r_\ast^\infty}^\infty \left( {Q-\omega^2 \over Q^{1/2} + \omega}
\right)  dr_\ast
\end{equation}
This is readily done by first expanding the integrand in inverse
powers of $r$ and then integrating. The result is convergent in
all relevant cases.


\subsection{Numerical implementation}

Before concluding the description of the methods we have devised
to study black hole resonances, it is worth commenting on  
the numerical implementation of the schemes. Overall we have a system of two coupled
nonlinear
first-order equations to solve numerically (for either choice of phase-function), 
namely, eqns (\ref{P1c}) or (\ref{P2c}) together with (\ref{tortc}). 
In reality, since we are dealing with intrinsically complex quantities, we 
split each equation in its real and imaginary parts. At the end of the day, we 
arrive at four equations to be solved. This is done by means of a standard 
variable-step Addams method.

To complete the solution of the problem, we combine the result of the integration 
of the phase functions with a 
complex root-finder routine (making use of M\"uller's algorithm \cite{recipes})
 to search the 
complex frequency/angular momentum plane 
for  zeros of the ratio  $A_{\rm in}/A_{\rm out}$.
A prior knowledge of the approximate 
location of a resonance greatly speeds up the overall procedure.  


\section{Numerical results}

The main aim of this investigation was to devise simple and efficient
methods for computing black-hole resonances and scattering phase-shifts. 
As should be clear from the previous Sections, the phase-function
schemes that we are discussing are straightforward to 
implement. They also turn out to be numerically robust 
and provide results of excellent accuracy. 
We have tested the methods by calculating known results for both
Schwarzschild and Kerr black holes. The obtained results serve as a 
validation of our numerical code(s). Having confirmed the 
reliability of our schemes we have extended the 
study to, in particular, the resonances of rotating black holes. 
Below, we will describe new results for both  
QNMs and Regge poles of Kerr black holes. These results shed interesting 
light on issues concerning the excitation of the long lived QNMs
that are known to exist for near extreme Kerr black holes, 
the relation between the approximate QNMs for $a \to M$ used in (for example)
\cite{kg1}, and also the effect of rotation on the 
black hole Regge poles.  


\subsection{Quasinormal modes}

As a first application of our numerical methods, we have 
calculated the first QNMs of a Schwarzschild black hole. 
The obtained results agree perfectly with results in the literature \cite{na2}, 
and confirm that our method is reliable as long as the modes are not 
damped too rapidly. The method breaks down when the imaginary part of the 
mode frequency is about one order of magnitude larger than the real part. 
This means that one should not expect to be able to use our scheme 
to study the asymptotic behaviour of the QNM spectrum. However, 
all astrophysically relevant QNMs are readily calculated. 
We have also tested the obtained excitation coefficients by 
comparing results obtained for gravitational perturbations
(using a code based on the Sasaki-Nakamura equation \cite{SN})
to the phase-integral results in \cite{naprd0} and the data obtained 
by Leaver \cite{leaver1}.
A sample of results is given in Table~\ref{tab1}. By comparing the listed results 
to those in Table~II of \cite{naprd0}  we conclude that the 
present results, which  agree very well with those  of Leaver apart from 
a few sign discrepancies already pointed out in \cite{naprd0}, are reliable.
In fact, we believe that the present method provides the (so far) 
most efficient tool for investigating the QNM excitation problem.
 
\begin{table}

\begin{tabular}{|c|c|c|r|} 
\hline
$l$ &  $n$  &$\omega_n  M$ & $B_n\qquad\qquad$ \\  
\hline
 2 &  0 &   $0.373672-0.088962i$ &  $0.126845+0.020669i$\\
   &  1 &   $0.346711-0.273915i$ &  $0.048328-0.223618i$\\
   &  2 &   $0.301052-0.478276i$ & $-0.190336+0.015118i$\\
   &  3 &   $0.251504-0.705148i$ &  $0.080519+0.079964i$\\
   &  4 &   $0.207514-0.946845i$ & $-0.016727-0.060630i$\\
   &  5 &   $0.169299-1.195608i$ & $-0.001566+0.035783i$\\
\hline
 3 & 0 &    $0.599443-0.092703i$ & $-0.093672-0.049606i$\\
   &  1 &   $0.582643-0.281298i$ & $-0.152350+0.269067i$\\
   &  2 &   $0.551683-0.479093i$ &  $0.414353+0.143038i$\\
   &  3 &   $0.511956-0.690338i$ & $-0.041146-0.412997i$\\
   &  4 &   $0.470173-0.915660i$ & $-0.220524+0.225274i$\\
\hline
 4 & 0 &    $0.809178-0.094164i$ &  $0.064924+0.065661i$\\
   & 1 &    $0.796631-0.284334i$ &  $0.263139-0.249790i$\\
   & 2 &    $0.772710-0.479904i$ & $-0.546231-0.439093i$\\
   & 3 &    $0.739835-0.683916i$ & $-0.323148+0.835511i$\\
   & 4 &    $0.701524-0.898240i$ &  $0.867222-0.114054i$ \\
\hline
\end{tabular}
\caption{Numerical results for the first few QNMs, and the associated 
excitation coefficients $B_n$, corresponding to 
gravitational perturbations of a Schwarzschild black hole.}
\label{tab1}
\end{table}

As the black hole acquires rotation, each  of the Schwarzschild QNMs (for a given $l$) 
splits into  $2l +1$ distinct modes.  We have tested the reliability of our scheme by 
reproducing the Kerr QNM frequencies obtained by Leaver, cf. Tables~2 and 3 
in \cite{leaver1}. A sample of results for $l=2$,  obtained by implementing our 
Pr\"ufer phase function scheme for the Sasaki-Nakamura equation, are provided in 
Table~\ref{tab2}. These results are in perfect agreement with those obtained by
Leaver's continued fraction calculation for $m=0$ and $m=\pm1$. 
Furthermore, Table~\ref{tab2} is useful since the $l=2$ QNMs may contribute 
significantly to the gravitational-wave signal from, for example, a black hole born 
following a supernova core collapse or a compact binary merger (see \cite{finn} for 
discussion).

\begin{table}
\begin{tabular}{|c|c|c|c|c|c|} 
\hline
$a/M$ & $m=2$ & $m=1$ & $m=0$ & $m=-1$ & $m=-2$ \\
\hline
0    & $0.373672-0.088962i$     & $0.373672-0.088962i$ & $0.373672-0.088962i$ & $0.373672-0.088962i$
& $0.373672-0.088962i$ \\
0.10 &     $0.387018-0.088706i$ & $0.380432-0.088798i$ & $0.374032-0.088898i$ & $0.367812-0.089004i$
& $0.361768-0.089114i$\\
0.20 &     $0.402145-0.088311i$ & $0.388248-0.088488i$ & $0.375124-0.088700i$ & $0.362738-0.088935i$
& $0.351053-0.089182i$\\
0.30 &     $0.419527-0.087729i$ & $0.397330-0.087995i$ & $0.376985-0.088353i$ & $0.358366-0.088763i$
& $0.341333-0.089184i$\\
0.40 &     $0.439842-0.086882i$ & $0.407979-0.087257i$ & $0.379682-0.087827i$ & $0.354633-0.088484i$
& $0.332458-0.089131i$\\
0.50 &     $0.464123-0.085639i$ & $0.420632-0.086173i$ & $0.383318-0.087069i$ & $0.351491-0.088091i$
& $0.324307-0.089031i$\\
0.60 &     $0.494045-0.083766i$ & $0.435969-0.084564i$ & $0.388054-0.085995i$ & $0.348911-0.087566i$
& $0.316784-0.088892i$\\
0.70 &     $0.532599-0.080794i$ & $0.455122-0.082085i$ & $0.394129-0.084453i$ & $0.346870-0.086884i$
& $0.309808-0.088717i$\\
0.80 &     $0.586015-0.075626i$ & $0.480231-0.077955i$ & $0.401917-0.082156i$ & $0.345356-0.086003i$
& $0.303313-0.088512i$\\
0.90 &     $0.671634-0.064863i$ & $0.516292-0.069804i$ & $0.412004-0.078483i$ & $0.344359-0.084865i$
& $0.297244-0.088281i$\\
0.99 &     $0.870861-0.029510i$ & $0.572720-0.046209i$ & $0.423685-0.072701i$ & $0.343890-0.083550i$
& $0.292107-0.088052i$\\
\hline\end{tabular}
\label{tab2}\caption{QNM frequencies ($\omega_n M$) 
for gravitational perturbations of Kerr black holes with various spin rates.
The data is for $l=2$, and illustrates that the co-rotating modes $m>0$ become long lived
as the black hole approaches extreme rotation. In contrast the counter-rotating modes
$m<0$ are much less affected by the rotation.  }
\end{table}

\begin{figure}[tbh]
\centerline{\epsfysize=6cm  \epsfbox{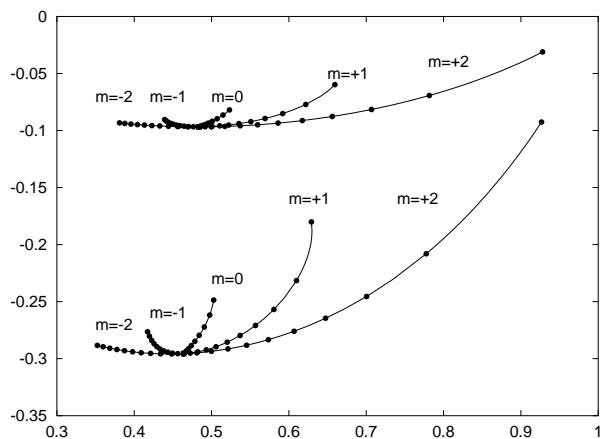}}
\caption{The trajectory (Im~$\omega M$ as a function of Re~$\omega M$) of 
the first two  $l=2$ scalar field Kerr QNMs for varying black hole rotation 
$ 0 \leq a/M \leq 0.99$. The position of each mode at spin values 
$a/M= 0, 0.1, 0.2 ... 0.9, 0.99$ is represented by dots.} 
\label{kerrq}
\end{figure}

Kerr QNMs are known to exhibit interesting behaviour as $a$ varies \cite{detweiler1}. 
In particular, we see from Table~\ref{tab2} that rotation has a pronounced
effect on modes with $m= l >0$. These can be thought of as ``equatorial''  
and ``co-rotating'' with the black hole (cf. the symmetry properties of the 
associated angular functions, which in the case of scalar waves  are spheroidal 
harmonics $S^{a\omega}_{lm}(\theta)$).  
In the limit $a\to M$ the co-rotating modes have been found to become long-lived
(Im~$\omega_nM \to 0$) and accumulate at the frequency $m\omega_{+}$ 
\cite{detweiler1,leaver2,onozawa1,kg1}.
Hence, rotation has a significant effect on these modes. 
In contrast, the ``counter-rotating'' modes ($ m < 0$) are much less affected by 
rotation and remain relatively 
close to the initial Schwarzschild QNMs for all values of $a$.
 In Fig.~\ref{kerrq} we illustrate this behaviour by showing
the change in the first few 
scalar field QNM frequencies as the rotation rate of 
the black hole is varied. The data in the figure corresponds to  
$ 0 \leq a/M \leq 0.99 $.

\begin{figure}[tbh]
\centerline{\epsfysize=6cm  \epsfbox{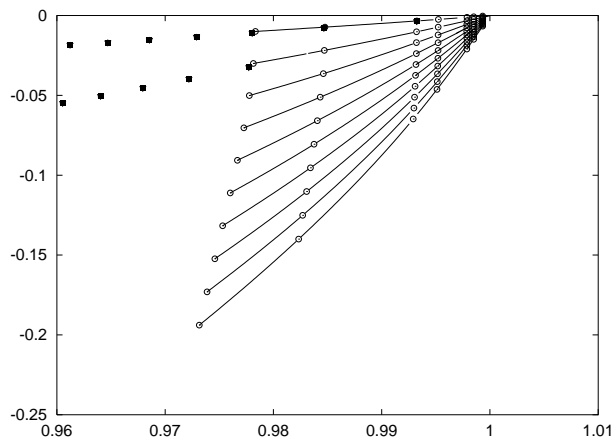}}
\caption{ Comparison of the numerical $l=m=2$ long-lived modes (the first
two are shown here, denoted by boxes) with modes derived by solving 
Detweiler's approximate mode-condition, (represented by lines, with the circles denoting 
the modes at $ a/M= 0.999, 0.9995, ... , 0.999999 $). It is clear that, for 
the fundamental mode, the two calculations are in agreement. 
The same seems to be true for the first overtone, although we have not been able to 
proceed numerically beyond $ a=0.999M$ in that case. The situation for the higher 
overtones also remain unclear.} 
\label{llmodes}
\end{figure}

The behaviour of the QNMs as the black hole approaches extreme rotation is 
particularly intriguing.
Unfortunately, our method breaks down beyond $a\approx 0.999M$. The likely reason 
for this is the multivaluedness of the complex $r_\ast$ coordinate, cf. for example 
(\ref{singu}), and the  need to initiate the integration close  
to the horizon, see eqn (\ref{spiral}). This becomes tricky as the 
event horizon and the inner Cauchy horizon come close together. 
A similar problem 
is known to arise in the phase-integral study of the QNMs of a 
Reissner-Nordstr\"{o}m black hole \cite{na6}. 
This means that we cannot push our calculation all the way to 
the  extreme limit. Of course, 
we can determine QNMs for all rotation rates that are likely to be astrophysically 
relevant. 
Still, the behaviour in the extreme limit is interesting from a conceptual point of
view. Especially since one can approximate the QNM spectrum for near extreme 
black holes \cite{detweiler1}. In figure~\ref{llmodes}
we compare our numerical results for scalar field QNMs 
to approximate results for  $a \approx M $ and 
$\omega \approx m\omega_{+}$. The figure shows that the numerical results
for the first two $l=m=2$ QNMs, as extended from the Schwarzschild limit, approach 
the first two approximate modes as $a\to M$. This is very encouraging since it
demonstrates a correspondence between two rather different computational schemes. 
It also provides  support for any argument concerning the nature and relevance of these 
long-lived modes that is based on the approximate formulae. 
The comparison also indicates the range of validity of Detweiler's approximate 
mode-condition \cite{detweiler1} in a useful way.

It has been argued that the existence of long-lived QNMs for 
rapidly rotating black holes could prove important for gravitational-wave 
astronomy. The basic idea is that a slowly damped mode will be easier to detect 
provided that it is initially excited to the same amplitude as a more rapidly 
damped mode. 
Since we now have the means to calculate the excitation coefficients for the 
Kerr QNMs we can analyse this problem quantitatively. As we have indicated in a previous 
paper \cite{kg1}, our results do not support the idea that the long-lived QNMs will be 
of particular importance. The reason for this can be seen from Table~\ref{tab2b} , 
where we provide results for the fundamental $m=\pm 2$ QNMs for various values of 
the rotation parameter. From this data we see that while the absolute value of the 
excitation coefficient for the counter-rotating QNM only changes by about a factor of 
two as the black hole spins up towards the extreme limit, it drops by about two order of 
magnitude for the co-rotating mode. This indicates that the co-rotating long-lived QNMs 
may be significantly more difficult to excite in a dynamical process. 

\begin{table}

\begin{tabular}{|c|cr|cr|} 
\hline
$a/M$ & \multicolumn{2}{c|}{$m=+2$} & \multicolumn{2}{c}{$m=-2$} \\
\hline   
 & $\omega_n M$ & $B_n\quad \qquad$ & $\omega_n M$ & $B_n\quad \qquad$ \\  
\hline
0.0   & $0.483644-0.096759i$ & $ 0.5252+0.0950i$  & $0.483644-0.096759i$ & $0.5252+0.0950i$ \\
0.1   & $0.499482-0.096666i$ & $ 0.6708+0.2710i$  & $0.469295-0.096708i$ & $0.4354+0.0358i$\\       
0.2   & $0.517121-0.096382i$ & $ 0.8355+0.8432i$  & $0.456198-0.096546i$ & $0.3790+0.0132i$\\    
0.3   & $0.536979-0.095839i$ & $ 0.8370+1.8608i$  & $0.444168-0.096299i$ & $0.3412+0.0044i$\\
0.4   & $0.559647-0.094931i$ & $-0.9646-0.0827i$  & $0.433059-0.095984i$ & $0.3145+0.0016i$\\               
0.5   & $0.585990-0.093494i$ & $-0.3913-0.2772i$ & $0.422751-0.095616i$ &  $0.2950+0.0018i$\\
0.6   & $0.617364-0.091245i$ & $-0.1359-0.2436i$ & $0.413146-0.095204i$ &  $0.2803+0.0034i$\\
0.7   & $0.656099-0.087650i$ & $ 0.0005-0.1699i$ & $0.404163-0.094759i$ &  $0.2691+0.0057i$\\       
0.8   & $0.706823-0.081520i$ & $ 0.0638-0.0754i$ & $0.395734-0.094285i$ &  $0.2607+0.0083i$\\    
0.9   & $0.781638-0.069289i$ & $ 0.0432+0.0143i$ & $0.387799-0.093790i$ &  $0.2550+0.0108i$ \\
0.99  & $0.928026-0.031069i$ & $-0.0027-0.0030i$ & $0.381041-0.093330i$ &  $0.2545+0.0123i$\\    
\hline
\end{tabular}
\caption{QNM frequencies ($\omega_n M$) and associated excitation 
coefficients ($B_n$)
for scalar perturbations of Kerr black holes with various spin rates.
The data is for $l=2$ and $m=\pm2$.}
\label{tab2b}
\end{table}

\begin{figure}[tbh]
\centerline{\epsfysize=6cm \epsfbox{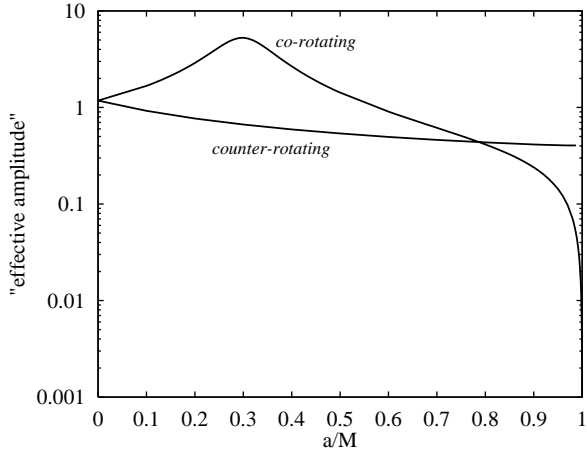}} 
\caption{ Assessment of the ``detectability'' of  Kerr 
black hole QNMs as $a\to M$.
The effective amplitude $h_{\rm eff}$ (as estimated from Eq. (\ref{heff2}))
is plotted as a function of the spin
parameter $a/M$. We compare the slowest damped co-rotating 
QNM to  the slowest damped counter-rotating one.  
The figure shows that the co-rotating mode (which becomes 
very long lived in the near extreme case) has a much smaller excitation 
coefficient than the counter-rotating mode as $a\to M$.}
\label{longfig}
\end{figure}

We can use the results in Table~\ref{tab2b} to discuss the 
``detectability'' of the long-lived QNMs.
Following \cite{kg1} we express the effective gravitational-wave amplitude 
associated with a single QNM (with complex frequency $\omega_n$) as
\begin{equation}
h_{\rm eff} \propto \sqrt{ { \mbox{Re } \omega_n  
\over \mbox{Im }
\omega_n } } {A^{\rm out}\over \alpha_n} = 2 \sqrt{ { \mbox{Re } \omega_n  
\over \mbox{Im }
\omega_n } } \omega_n B_n
\label{heff2}
\end{equation}
Given our numerical data, we can compare the result for co- and counter-rotating 
(scalar field) modes. Figure~\ref{longfig} (which is identical to Figure~1 
in \cite{kg1})  
shows the outcome of this comparison. From the figure it is clear that, 
even though the co-rotating mode
is much longer lived, its ``detectability'' decreases
significantly as $a\to M$. We conclude that the present results 
suggest that the effective amplitude of the slowest damped QNM for a 
rapidly rotating black hole may be several orders of magnitude 
smaller than the amplitude 
of the corresponding mode of a Schwarzschild black hole, or indeed the counter-rotating 
Kerr modes. The possible constructive interference of the many long-lived QNMs
that exist for near extreme Kerr black holes, and which could still lead to a 
relevant gravitational-wave signal, has been discussed in \cite{kg1}.

\begin{figure}[tbh]
\centerline{\epsfysize=6cm  \epsfbox{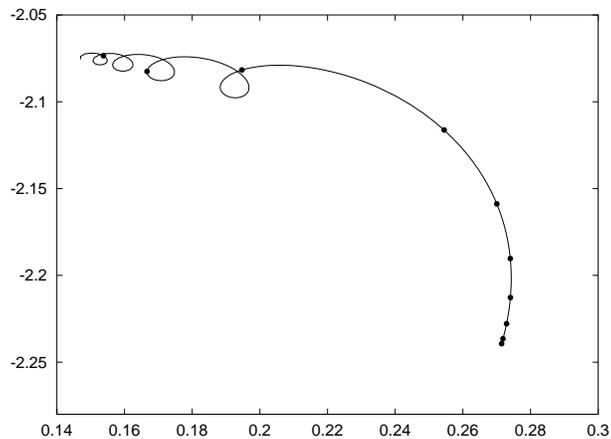}}
\caption{The trajectory of the ninth $\ell=2, m=0$ overtone for 
$ 0 \leq a/M \leq 0.92 $ (we show Im~$\omega_n M$ as a function of Re~$\omega_n M$).
Dots represent the position of the mode for $ a/M= 0, 0.1, ... 0.9 $. The
characteristic ``looping'' motion is prominent for moderate to high spins.} 
\label{overt}
\end{figure}

As a final application of our method we have traced the path of a high overtone
mode ($n=9, l=2, m=0$) for varying black-hole spin. The result is shown 
in Fig.~\ref{overt}. 
The purpose of this exercise was to investigate whether the
QNM trajectory exhibits the characteristic ``looping'' behaviour, first 
found in the studies of gravitational and electromagnetic field QNMs for Kerr and 
Reissner-Nordstr\"{o}m black holes \cite{onozawa1,onozawa2}.
 Fig.~\ref{overt} confirms that this  behaviour is also present for scalar fields.
It has been suggested \cite{onozawa1,onozawa2} that this peculiar behaviour is 
somehow associated with the change in the potential near to the horizon as the 
spin varies. However, since it is unclear how the high overtone modes are  
related to the potential a fully satisfactory explanation 
is still missing (see, however, \cite{liu} for an interesting point of 
view.

\subsection{Regge poles}

As mentioned in the Introduction, Regge pole calculations are
comparatively easy to perform  since as the associated wavefunction
(\ref{in}) remains well behaved as $r_{\ast} \to \pm \infty$. 
This means that one does not have to deal with divergences, and the 
identification of in- and outgoing wave solutions is straightforward.
For a given complex $l$, the set of equations (\ref{P1}) and (\ref{P2}) 
can be directly integrated along the real $r_{\ast}$-axis without any serious problems. 

To test our numerical code
we have calculated the first few Regge poles for a Schwarzschild black
hole for three different frequencies. The obtained results are listed in
Table~\ref{tab3}. As far as the Regge pole positions are concerned,
the results  are in agreement with the approximate phase-integral results 
given by Andersson \cite{cam2}. However, our calculation has confirmed
an earlier suggestion by Jensen \cite{jensen3} that the phase-integral residues
had a systematic error. As in the case of QNMs, we believe the present
method is the most efficient and reliable way to calculate black-hole Regge
poles.
 
\begin{table}
\begin{tabular}{|c|c|c|} 
\hline
 $\omega M$  & $l_n$ & $r_n$ \\
\hline
   $0.500$   & $2.086844+0.504735i$  & $0.527566+0.019735i$ \\
             & $2.188325+1.479613i$  & $0.448169-0.888761i$\\
             & $2.342582+2.392097i$  &$-0.350846-1.171626i$\\
\hline
   $1.000$   & $4.690075+0.501236i$  & $0.473472+0.575418i$\\
             & $4.742988+1.494954i$  & $2.369138-1.190595i$\\
             & $4.839323+2.466091i$  & $0.045487-5.147693i$\\
\hline
   $2.000$   & $9.889215+0.500325i$  &$-0.879437+0.579492i$\\
             & $9.915868+1.498744i$ &  $3.340145+6.574819i$\\
             & $9.967248+2.490034i$ & $25.613332-5.812667i$ \\
\hline
\end{tabular}
\caption{The first three Schwarzschild Regge poles ($l_n$) 
and the associated residues ($r_n$) for three 
different (real) frequencies $\omega M$.}
\label{tab3}
\end{table}

\begin{figure}[tbh]
\centerline{\epsfysize=7cm  \epsfbox{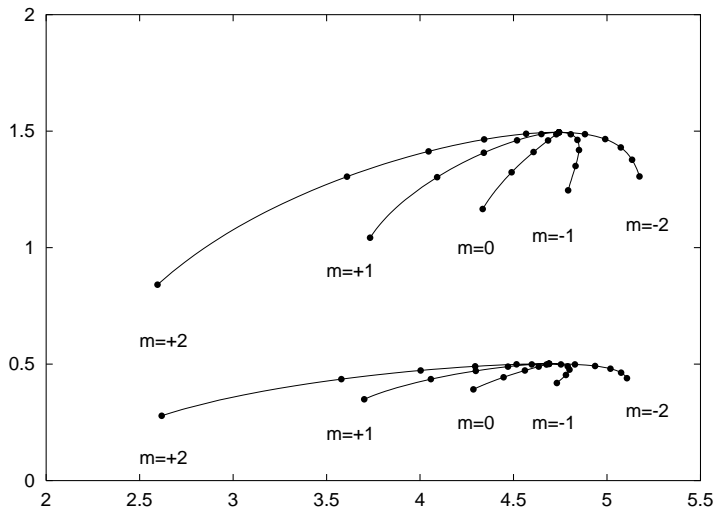}}
\caption{The first two Kerr Regge poles (we show Im~$l_n$ as a function
of Re~$l_n$) for various black hole rotation rates $ 0 \leq a/M \leq 0.99 $. Dots
represent the position of each pole for $a/M= 0, 0.2, 0.4, ..., 0.99$.
The frequency is $\omega M=1$. Note the distinct difference between the 
$m > 0$ and $m <0 $ branches which suggests that the former (latter) can be associated 
with prograde (retrograde) propagating waves. There are also obvious similarities in 
the behaviour of Regge poles and QNMs (compare this figure with Fig.~\ref{kerrq})} 
\label{reggK}
\end{figure}

We next consider the Regge poles of a Kerr black hole. 
This is a problem that, as far as we are aware, has not been discussed before. 
As in the case of QNMs, the loss of spherical symmetry 
in the background spacetime unfolds the $2l + 1$ degeneracy of the 
non-rotating resonances. The pole trajectories 
(for the first two Regge poles) are illustrated in Fig.~\ref{reggK} for 
 $\omega M=1$ and $ 0 \le a/M  \le 0.99 $, and a sample of numerical data 
is given in Table~\ref{tab4}.        
Figure~\ref{reggK} shows obvious similarities with the QNMs trajectories
in figure~\ref{kerrq}. This is not surprising given the 
close relationship between the two classes of scattering resonances \cite{jensen1}.

\begin{table}
\begin{tabular}{|c|c|c|} 
\hline 
 $a/M$ &$m=+2$ & $m=-2$ \\
\hline
0.0   & $4.690075+0.501235i$ & $4.690075+0.501235i$  \\
0.1   & $4.608103+0.500711i$ & $4.763131+0.500574i$  \\       
0.2   & $4.516055+0.498888i$ & $4.828131+0.498790i$  \\    
0.3   & $4.412352+0.495582i$ & $4.885722+0.495906i$  \\
0.4   & $4.294773+0.490504i$ & $4.936387+0.491909i$  \\                                                                    
0.5   & $4.160054+0.483194i$ & $4.980478+0.486748i$  \\
0.6   & $4.003103+0.472870i$ & $5.018235+0.480332i$  \\
0.7   & $3.815191+0.458088i$ & $5.049802+0.472521i$  \\       
0.8   & $3.578931+0.435695i$ & $5.075234+0.463109i$  \\    
0.9   & $3.247925+0.396177i$ & $5.094504+0.451794i$  \\
0.99  & $2.617843+0.278170i$ & $5.106494+0.439626i$  \\                                                      
\hline
\end{tabular}
\caption{Numerical results for the fundamental $m= \pm 2$  Kerr Regge poles 
($l_n$) for a selection of black hole spins. The frequency is $\omega M=1$.}
\label{tab4}
\end{table}

The most important conclusion we can draw from our results 
concerns the physical interpretation of the Kerr Regge poles.
First of all, it is clear that $m >0$ $(m<0)$ Regge poles would 
describe co-rotating (counter-rotating) waves. 
Intuitively, we would expect that the 
 ``angular life'' (the inverse of Im~$l_n$, cf. \cite{cam2})
of the former would be 
significantly increased due to rotational frame dragging. As in 
the case of QNMs, we find that the $m=l$ poles are the most ``sensitive'' 
to the black hole's rotation. This strongly suggests that these poles can be 
associated with waves propagating around the black hole 
in the vicinity of the equatorial plane. 
More evidence to support this intuitive picture comes from the behaviour
of Re~$l_n$. The decrease (increase) of this quantity, as the black hole 
spins up, is qualitatively expected from the corresponding decrease (increase)
of the critical impact parameter and unstable photon orbit for prograde
(retrograde) motion in Kerr spacetime. 

\section{Concluding remarks}

We have developed simple numerical methods 
for calculating black hole resonances (QNMs and Regge poles). 
The basic ingredients of our schemes are (i) the use of (Pr\"{u}fer) phase functions 
in place of the original wavefunction (ii) 
the integration of the resulting equations along rotated contours 
in the $r_{\ast}$-plane (when the frequency is complex), and (iii) the 
implementation of M\"uller's algorithm for locating the complex roots
of $1/{\cal S}(\omega, l)$ function 
(the inverse scattering matrix element, whose zeros define the resonances).   

We have applied our schemes to the problem of calculating gravitational and
massless scalar field perturbation 
QNMs and Regge poles for rotating and non-rotating black holes. 
The obtained numerical data verify all existing results in the literature
with high precision. Moreover, we have provided some new results. 
In particular, we have shown how the slowest damped Kerr QNMs 
match the  ``long-lived'' modes of a near extreme black hole 
which can be derived by an analytic approximation to the Teukolsky equation. 
This ties up some loose ends as far as the Kerr QNMs are concerned and 
also provides important support for using the approximate results
to discuss the dynamics of near extreme black holes \cite{kg1}. 
Our results for the excitation coefficients of Kerr QNMs add fuel 
to the debate concerning the detectability of gravitational 
waves associated with the long lived modes of a near extreme 
black hole. We have also provided the first results for
Kerr Regge poles and discussed briefly how these results make 
sense given the established physical interpretation of their Schwarzschild 
counterparts. 

In conclusion, we find that our methods can be regarded as reliable and useful, 
apart from (i) for very rapidly damped QNMs, and (ii) in the case of near extreme 
rotations ($a/M>0.999$ or so). This is, however, a small price to pay considering 
the profound simplicity and adaptability of the method. This is the main advantage 
of our approach. All other accurate schemes require a more sophisticated analysis: 
Leaver's continued fraction method \cite{leaver1} is not easy to implement given 
the slow convergence of the relevant series, and the complex-coordinate 
phase-amplitude method developed by one of us \cite{na2} makes use of much 
more intricate integration contours than we used in the present analysis. 
Finally, we should note that, even though we considered mainly 
the case of scalar perturbations in this paper, our method can readily be used 
for the more realistic case of gravitational perturbations. 
We therefore believe that our ``quick and dirty'' approach to black hole
resonances may prove useful in a variety of exciting 
contexts.

\acknowledgements

The work of K.G. was financially supported by the State Scholarships Foundation 
of Greece. N.A. is a Philip Leverhulme Prize Fellow and also acknowledges support 
from the EU Programme 'Improving the Human Research Potential and the Socio-Economic
Knowledge Base' (Research Training Network Contract HPRN-CT-2000-00137).



\begin{references}

\bibitem{novikov} 
        N. Andersson, Chapter 4 in {\em Black-hole physics } by V.P. Frolov and
        I.D. Novikov (Kluwer, Dordrecht 1998)

\bibitem{vishu}  
        C.V. Vishveshwara, {\em Nature} {\bf 227}, 936 (1969)


\bibitem{kokkotas}
        K.D. Kokkotas and B.G. Schmidt, {\em Living Reviews in Relativity},
        electronic journal http://www.livingreviews.org (1999)

\bibitem{nollert_rev}
        H.-P. Nollert, {\em Class. Quantum Grav.} {\bf 16}, R159 (1999)

         
\bibitem{chandra1}
        S. Chandrasekhar and S. Detweiler, {\em Proc. Roy. Soc. (London) A}
        {\bf 343 }, 289 (1975)


\bibitem{detweiler1} 
        S. Detweiler, {\em Astrophys.\  J}.\  {\bf 225}, 687 (1978)


\bibitem{valeria}
         V. Ferrari and B. Mashhoon, {\em Phys. Rev. D} {\bf 30}, 295, (1984)


\bibitem{schutz}
        B.F. Schutz and C.M. Will, {\em Astrophys. J.} {\bf 291}, L33 (1985)

\bibitem{iyer1}
         S. Iyer and C.M. Will, {\em Phys. Rev. D} {\bf 35}, 3621 (1987) 


\bibitem{iyer2}
         S. Iyer, {\em Phys. Rev. D} {\bf 35}, 3632 (1987)

\bibitem{iyer3}
         E. Seidel and S. Iyer, {\em Phys. Rev. D} {\bf 41}, 374 (1990)


\bibitem{guinn}
        J.W. Guinn, C.M. Will, Y. Kojima and B.F. Schutz, {\em Class. Quantum
        Grav. } {\bf 7}, L47 (1990) 

\bibitem{leaver1}
         E.W. Leaver, {\em Proc. R. Soc. London A} {\bf 402}, 285 (1985)  


\bibitem{leaver2}
        E.W. Leaver, {\em Phys. Rev. D}  {\bf 34}, 384 (1986)


\bibitem{na2}
        N. Andersson, {\em Proc. R. Soc. London  A} {\bf 439}, 47 (1992)

\bibitem{nollert1}
       H.-P. Nollert and B.G. Schmidt, {\em Phys. Rev. D} {\bf 45}, 2617 (1992)


\bibitem{berry}
        M. Berry and K.E. Mount, {\em Rep. Prog. Phys. } {\bf 35}, 315 (1972)


\bibitem{froman} 
        N. Fr\"{o}man and P.O. Fr\"{o}man, 
        {\em JWKB approximation: Contributions to the Theory }, 
        (Amsterdam: North Holland 1965)


\bibitem{pi}
     N.Andersson and S. Linn\ae us, {\em Phys. Rev. D} {\bf 46}, 4179 (1992);
     N. Fr\"{o}man, P.O. Fr\"{o}man, N. Andersson and A. H\"{o}kback,
     {\em Phys. Rev. D} {\bf 45}, 2609 (1991); 
     N. Andersson, M.E. Ara\'{u}jo and B.F. Schutz, {\em Class. Quantum Grav.}
     {\bf 10}, 757 (1993);
     

\bibitem{onozawa1}
         H. Onozawa,  {\em Phys. Rev. D} {\bf 55}, 3593 (1997)

\bibitem{kg1}
        K. Glampedakis and N. Andersson, {\em Phys. Rev. D} {\bf 64}, 104021 (2001) 

\bibitem{press}
      S.A. Teukolsky, W.H. Press, {\em Astrophys. J.} {\bf 193}, 443 (1974)

\bibitem{bk03} E. Berti and K.D. Kokkotas,  {\em Asymptotic quasinormal modes of Reissner-Nordstr\"om and Kerr black holes}, preprint hep-th/0303029

\bibitem{noll} H.-P. Nollert, {\em Phys. Rev. D} {\bf 47}, 5253 (1993) 


\bibitem{nahigh} N. Andersson, {\em Class. Quantum Grav.} {\bf 10}, L61 (1993)


\bibitem{dreyer} O. Dreyer, {\em Quasinormal Modes, the Area Spectrum, and Black Hole Entropy} preprint gr-qc/gr-qc/0211076


\bibitem{motl} L. Motl, {\em An analytical computation of asymptotic Schwarzschild quasinormal frequencies}, preprint gr-qc/0212096

\bibitem{hod} S. Hod, {\em Phys. Rev. Lett.} {\bf 81}, 4293 (1998)

\bibitem{jensen1} Y. D\'ecanini, A. Folacci and B.P. Jensen, {\em Complex 
angular momentum in black hole physics and the quasi-normal modes} preprint gr-qc/0212093

\bibitem{nussenzveig}
        H.M. Nussenzveig, {\em Diffraction effects in Semiclassical
        Scattering} (Cambridge: Cambridge University Press 1992)


\bibitem{cam1}
        N. Andersson and K-E. Thylme, {\em Class. Quantum Grav.} {\bf 11},
        2991 (1994)

\bibitem{cam2} 
        N.Andersson, {\em Class. Quantum Grav.} {\bf 11}, 3003 (1994)

\bibitem{chandra2}
       S. Chandrasekhar and V. Ferrari, {\em Proc. R. Soc. London A} 
      {\bf 437}, 133 (1992)

\bibitem{kg2}
        K. Glampedakis and N. Andersson, {\em Class. Quantum Grav.} {\bf 18}, 
        1939 (2001)

\bibitem{taylor}
        J.R. Taylor, {\em Scattering Theory} (New York:Willey, 1972)

\bibitem{chandrabook}
       S. Chandrasekhar, {\em The Mathematical Theory of Black Holes}
       (Oxford University Press, New York, 1983)

\bibitem{teuk}
      S.A. Teukolsky, {\em Astrophys. J.} {\bf 185}, 635 (1973)

\bibitem{jensen2}
        B.P. Jensen, {\em Annales De Physique} {\bf 16}, 117 (1989)

\bibitem{lax}
        P.D. Lax and R.S. Phillips, {\em Scattering Theory} (Academic,
        New York, 1967)


\bibitem{naprd} N. Andersson, {\em Phys. Rev. D} {\bf 55} 468 (1997) 

\bibitem{matzner}
         R.A. Matzner, {\em J. Math. Phys. } {\bf 9}, 163 (1968)

\bibitem{handler}
        F.A. Handler and R.A. Matzner, {\em Phys. Rev D} {\bf 22}, 2331 (1980)

\bibitem{milne}
        W.E. Milne, {\em Phys. Rev. } {\bf 35}, 863 (1930)

\bibitem{newman}
       W.I. Newman and W.R. Thorson, {\em Phys. Rev. Lett.} {\bf 29}, 1350 (1972);
      {\em Can. J. Phys. } {\bf 50}, 2997 (1972)

\bibitem{pryce}
         J.D.Pryce, \textit{Numerical Solution of Sturm-Liouville Problems}
         (Oxford Science Publications, Oxford, England, 1993)

\bibitem{pajunen1}
         P. Pajunen, {\em J. Chem. Phys.} {\bf 88}, 4268, (1988)

\bibitem{pajunen2}
         P. Pajunen, {\em J. Comp. Phys.} {\bf 82}, 16, (1989)

\bibitem{cam3}
        N. Andersson, {\em J. Phys. A: Math. Gen. } {\bf 26}, 5085 (1993)

\bibitem{na4}
        N. Andersson, K.D. Kokkotas and B.F. Schutz, {\em Mon. Not. R. Astron.
         Soc.} {\bf 274}, 1039 (1995)

\bibitem{araujo}
        M.E. Ara\'{u}jo, D. Nicholson and B.F. Schutz, 
        {\em Class. Quantum Grav.} {\bf 10}, 1127 (1993)

\bibitem{recipes}
        W.H.Press, S.A. Teukolsky, W.T. Vetterling and B.P. Flannery, 
        \textit{Numerical Recipes} (Cambridge University Press, Cambridge,     
         England, 1992)

\bibitem{SN}
        M. Sasaki and T. Nakamura. {\em Prog. Theor. Phys. } {\bf 67},
        1788 (1982)

\bibitem{naprd0} N. Andersson {\em Phys. Rev. D} {\bf 51} 353 (1995). 

\bibitem{finn} L.S. Finn {\em Phys. Rev. D} {\bf 46 } 5236 (1992). 

\bibitem{na6}
        N. Andersson, {\em Proc. R. Soc. London  A} {\bf 442}, 427 (1993) 

\bibitem{onozawa2}
         N. Andersson and H. Onozawa, {\em Phys. Rev. D} {\bf 54}, 7470 (1996)


\bibitem{liu}
        H. Liu, {\em Class. Quantum Grav.} {\bf 12}, 543 (1995)


\bibitem{jensen3} B.P. Jensen, private communication

\end{references}
\end{document}